\documentclass[12pt]{iopart}
\usepackage{dcolumn}% Align table columns on decimal point
\usepackage{bm}% bold math
\usepackage{graphicx}
\usepackage{subfigure}
\usepackage[sort&compress,numbers]{natbib}
 
\begin{document}
\title{Dynamics of Confident Voting}
\author{D. Volovik$^1$ and S. Redner$^1$}
\address{$^1$Center for Polymer Studies and Department of
  	Physics, Boston University, Boston, Massachusetts 02215, USA}

\begin{abstract}

  We introduce the confident voter model, in which each voter can be in one
  of two opinions and can additionally have two levels of commitment to an
  opinion --- confident and unsure.  Upon interacting with an agent of a
  different opinion, a confident voter becomes less committed, or
  unsure, but does not change opinion.  However, an unsure agent
  changes opinion by interacting with an agent of a different opinion.  In
  the mean-field limit, a population of size $N$ is quickly driven to a mixed
  state and remains close to this state before consensus is eventually
  achieved in a time of the order of $\ln N$.  In two dimensions, the
  distribution of consensus times is characterized by two distinct times ---
  one that scales linearly with $N$ and another that appears to scale as
  $N^{3/2}$.  The longer time arises from configurations that fall into
  long-lived states that consist of two (or more) single-opinion stripes
  before consensus is reached.  These stripe states arise from an effective
  surface tension between domains of different opinions.

\end{abstract}

\maketitle

\section{INTRODUCTION}

The voter model~\cite{Liggett85} describes the evolution toward consensus in
a population of $N$ agents, each of which can be in one of two possible
opinion states.  In an update event, a randomly-selected voter adopts the
state of a randomly-selected neighbor.  As a result of repeated update
events, a finite population necessarily reaches consensus in a time $T_N$
that scales as a power law in $N$ (with a logarithmic correction in two
dimensions)~\cite{Liggett85,K92}.  Because of its simplicity and its natural
connection to opinion dynamics, the voter model has been extensively
investigated (see, e.g.,~\cite{CFL90,KRB10}).  The connection with social
phenomena has also motivated efforts to extend the voter model to incorporate
various aspects of social reality, such as, among others,
stubbornness/contrarianism~\cite{G04,GF07,MPR07,YAOSS11}, multiple
states~\cite{VR04,VMR09}, internal dissonance~\cite{PSS07}, individual
heterogeneity~\cite{MGR10}, environmental
heterogeneity~\cite{SR05,SES05a,SES05b,SAR08}, vacillation~\cite{LR07}, and
non-linear interactions~\cite{LR08,CMP09}.  These studies have uncovered many
new phenomena that are still being actively explored.

Our investigation was initially motivated by recent social experiments of
Centola~\cite{C10}, who studied the spread of a specific behavior in a
controlled online network where {\em reinforcement\/} played a crucial role.
Reinforcement means that an individual adopts a particular state only after
receiving multiple prompts to adopt this behavior from socially-connected
neighbors.  These experiments found that social reinforcement played a
decisive role in determining how a new behavior is adopted~\cite{C10}.

Previous research that has a connection with this type of reinforcement
mechanism include the q-voter model~\cite{CMP09}, where multiple same-opinion
neighbors initiate change, the naming game, and the AB model~\cite{CBL09}.
An example that is perhaps most closely connected to reinforcement arises in
the noise-reduced voter model~\cite{DC07}, where a voter keeps a running
total of inputs towards changing opinions, but actually changes opinions only
when this counter reaches a predefined threshold.  A similar notion of
reinforcement arises in a model of fad and innovation dynamics~\cite{KRV11}
and in a model of contagion spread~\cite{MWGP11}.  The use of multiple
discrete opinions is not the only option for incorporating varying opinion
strength.  Previous models have used a continuous range of opinions
quantifying the tendency for an agent to change its opinion.~\cite{M12} For
example, in the bounded confidence model, an agent can possesses an opinion
in a continuous range, with the spatial distance between points representing
the difference in those opinions.~\cite{FLPR05}

\begin{figure}[ht]
\centerline{\includegraphics*[width=0.75\textwidth]{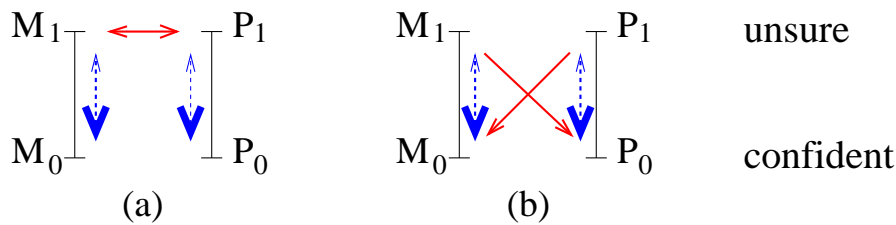}}
\caption{\small Illustration of the states and possible transitions in the:
  (a) marginal, and (b) extremal versions of the confident voter model.
  Dashed arrows indicate possible confidence level changes (biased toward
  higher confidence), while solid arrows indicate possible opinion change
  events.}
  \label{model}
\end{figure}

In this paper, we study how reinforcement affects the dynamics of the voter
model.  In our\emph{confident voter model\/}, we assume that agents possess
some modicum of intrinsic confidence in their beliefs and, unlike the classic
voter model, need multiple prompts before changing their opinion state.  We
investigate a simple realization of this confident voting in which each
opinion state is further demarcated into two substates of different
confidence levels.  The basic variables are thus the opinion of each voter
and the confidence level with which this opinion is held.  For concreteness,
we label the two opinion states as plus (P) and minus (M).  Thus the possible
states of an agent are $P_0$ and $P_1$ for confident and unsure plus agents,
respectively, and correspondingly $M_0$ and $M_1$ for minus agents
(Fig.~\ref{model}).  The new feature of confident voting is that a confident
agent does not change opinion by interacting with an agent of a different
opinion.  Instead such an agent changes from being confident to being unsure
of his opinion.  On the other hand, an unsure agent changes opinion by
interacting with any agent of the other opinion, as in the classic voter
model.

We define two variants of confident voting that accord with common anecdotal
experience (Fig.~\ref{model}).  In the \emph{marginal\/} version, an unsure
agent that changes opinion still remains unsure.  Such an agent is often
labelled a ``flip-flopper'', a routinely-invoked moniker by American
politicians to characterize political opponents.  Figuratively, an agent who
switches opinion remains ambivalent about the new opinion state and can
switch back.  In the \emph{extremal\/} version, an unsure agent becomes
confident after an opinion change.  Such an agent ``sees the light'' and
therefore becomes fully committed to the new opinion state.  This behavior is
typified by Paul the Apostle, who switched from being dedicated to finding
and persecuting early Christians to embracing Christianity after experiencing
a vision of Jesus.

\section{MEAN-FIELD DESCRIPTION}
\label{MF}

The basic variables are the densities of the four types of agents.  We use
$P_0, P_1, M_0, M_1$ to denote both the agent types and their densities.  In
the mean-field description, a pair of agents is randomly selected, and the
state of one the two agents, chosen equiprobably, changes according to the
voter-like dynamics illustrated in Fig.~\ref{model}.  We now outline the time
evolution for the two variations of the confident voter model.

\subsection{Marginal Version}

For writing the rate equations, we first enumerate the possible outcomes when
a pair of agents interact:
\begin{itemize}
\itemsep -0.5ex
\item[]\qquad $M_1P_1$  $\to $ $M_1M_1$ or $P_1P_1$;\qquad\qquad\qquad
$M_0P_0$\,\,  $\to$ $M_0P_1$\, or $M_1P_0$;
\item[]\qquad $P_0\,P_1$\,  $\to $ $P_0\,P_1$\,\,\, or $P_0P_0$;\qquad\qquad\qquad
$M_0M_1$  $\to $ $M_0M_1$ or $M_0M_0$;
\item[]\qquad $M_1P_0$  $\to$ $M_1P_1$\,\, or $P_0P_1$;\qquad\qquad\qquad
$M_0P_1$\,\,\,  $\to$ $M_1P_1$ \,or $M_0M_1$.
\end{itemize}
That is, the interaction between two unsure agents of opposite opinions
($M_1P_1$) leads to no \emph{net\/} density change, as in the classic voter
model.  However, when two confident agents of different opinions meet
($M_0\,P_0$), one of the agents becomes unsure.  The next two lines
account for interactions between agents of the same opinion but different
confidence levels.  We assume that an unsure agent exerts no influence on
a confident agent by virtue of the latter being confident, while a confident
agent is persuasive and converts an unsure agent to confident.  Finally,
the last line accounts for an unsure agent changing opinion upon
interacting with a confident agent of a different opinion.

The corresponding rate equations are:
\begin{eqnarray}
  \label{REM}
\dot P_0&= -(M_0+M_1)P_0+P_0P_1 \equiv -MP_0+P_0P_1\,,\\\nonumber
\dot P_1&=  ~~MP_0-P_0P_1+(M_1P_0-M_0P_1)\,,\nonumber
\end{eqnarray}
with parallel equations for $M_0$ and $M_1$ that are obtained by
interchanging $M\leftrightarrow P$ in Eq.~(\ref{REM}).  The rate equation for
the total density of plus agents is
\begin{equation*}
\dot P =M_1P_0-M_0P_1,
\end{equation*}
and from the complementary equation for $\dot M$, it is evident that the
total density of agents is conserved, $\dot P+\dot M=0$.

\subsection{Extremal Version} 

For the extremal version, we again enumerate the possible outcomes when a
pair of agents interact. These are:
\begin{itemize}
\itemsep -0.5ex
\item[]\qquad $M_1\,P_1$ \, $\to\, $ $P_0P_1$ or $M_0M_1$;
\qquad\qquad\qquad $M_0\,P_0$ \, $\to$ $M_0P_1$ \, or $M_1P_0$;
\item[]\qquad $P_0\,P_1$\,\,\,\,\,  $\to $  $P_0P_1$ \,or $P_0P_0$;
\qquad\qquad\qquad \,\,$M_0\,M_1$ \, $\to $  $M_0M_1$ \,or $M_0M_0$;
\item[]\qquad $M_1\,P_0$ \, $\to$ $M_1P_1$ or $P_0P_0$;
\qquad\qquad\qquad \,\,$M_0\,P_1$ \,\, $\to$ $M_1P_1$ \,\,or $M_0M_0$.
\end{itemize}
The point of departure, compared to the marginal version, is that a voter is
now confident in its new opinion state upon changing opinion.  The rate
equations corresponding to these steps are:
\begin{eqnarray}
\label{REE}
\dot P_0&= -M_0P_0+ M_1P_1+P_0P_1\,,\\\nonumber
\dot P_1&= ~~ M_0P_0-M_1P_1-P_0P_1+(M_1P_0\!-\!M_0P_1)\,,\nonumber
%&~~~~~~~~~~~~~-\tfrac{1}{2}P_0P_1\,,
\end{eqnarray}
with parallel equations for $M_0$ and $M_1$.  The rate equation for the total
density of plus agents is the same as that for the marginal version, so that
again the total density of agents is manifestly conserved.

\subsection{Time Evolution}

For both variants of the confident voter model, the time evolution is
dominated by the presence of a saddle point that corresponds not to
consensus, but a balance between plus and minus agents.  For nearly-symmetric
initial conditions, the densities of the different species are initially
attracted to this unstable fixed point, but eventually flow to a stable fixed
point that corresponds to consensus.  However when the initial condition is
perfectly symmetric between plus and minus agents, then the population is
driven to a mixed state that corresponds to the symmetric saddle point
(Fig.~\ref{non-symm}).

\subsubsection{Symmetric System}

It is instructive to first study the initial conditions $M_0(0)=P_0(0)=1/2$
and $M_1(0)=P_1(0)=0$.  The rate equations (\ref{REM}) for the marginal
version of confident voting now reduce to $\dot P_0=-\dot P_1=-P_0^2$, with
solution
\begin{eqnarray}
P_0(t)&=P_0(0)/[1+P_0(0)t]\,,\\\nonumber
P_1(t)&=\frac{1}{2}-P_0(t)\,.
\end{eqnarray}
Thus in an initially symmetric system, confident voters are slowly
eliminated because there is no mechanism for their replenishment, and all
that remains asymptotically are equal densities of unsure voters.

For the extremal version of confident voting, the rate equations (\ref{REE})
reduce to
\begin{eqnarray}
\label{dotP0}
\dot P_0&=-\dot P_1=P_0^2+\frac{1}{2}P_0-\frac{1}{4}\\\nonumber
&= -(P_0-\lambda_+)(P_0-\lambda_-)\,,
\end{eqnarray}
with $\lambda_\pm=\frac{1}{4}(-1\pm\sqrt{5})\approx 0.309, -0.809$.  Because
the quadratic polynomial on the right-hand side of Eq.~(\ref{dotP0}) is
positive for $P_0<\lambda_+$ and negative for $P_0>\lambda_+$, the fixed
point at $\lambda_+$ is stable.  Thus $P_0(t)$ approaches $\lambda_+$
exponentially in time.  We solve for $P_0$ by a partial fraction expansion to
give
\begin{equation}
\frac{P_0(t)-\lambda_+}{P_0(t)-\lambda_-} =
  \frac{P_0(0)-\lambda_+}{P_0(0)-\lambda_-}\,\,
e^{-(\lambda_+-\lambda_-)t}~,
\end{equation}
which indeed gives an exponential approach to the final state of
$P_0=\frac{1}{2}-P_1=\lambda_+$.  Thus all four voting states are represented
in the long-time limit.

\subsubsection{Non-Symmetric System}

If the initial condition is slightly non-symmetric, then numerical
integrations of the rate equations clearly show that the evolution of the
densities turns out to be controlled by two distinct time scales --- a fast
time scale that is $\mathcal{O}(1)$ and a longer time scale that is
$\mathcal{O}(\ln N)$, where $N$ is the population size.  To incorporate $N$
in the rate equations, we interpret these equations as describing the
dynamics of voters that live on a complete graph of $N\gg1$ sites, so that
every agent interacts equiprobably with any other agent.  In this framework,
consensus on the complete graph should be viewed as the density of a single
species being equal to $1-\frac{1}{N}$ in the rate equations.  Similarly, an
initial small deviation $\epsilon=\frac{1}{N}$ from the symmetric initial
conditions in the rate equations (i.e., $P_0(0)=\frac{1}{2}+\epsilon$ and
$M_0(0)=\frac{1}{2}-\epsilon$, with $\epsilon=\frac{1}{N}$), should be
interpreted as the departure from a symmetric state by a single particle on a
complete graph of $N$ sites.

\begin{figure}[ht]
\centerline{\includegraphics*[width=0.425\textwidth]{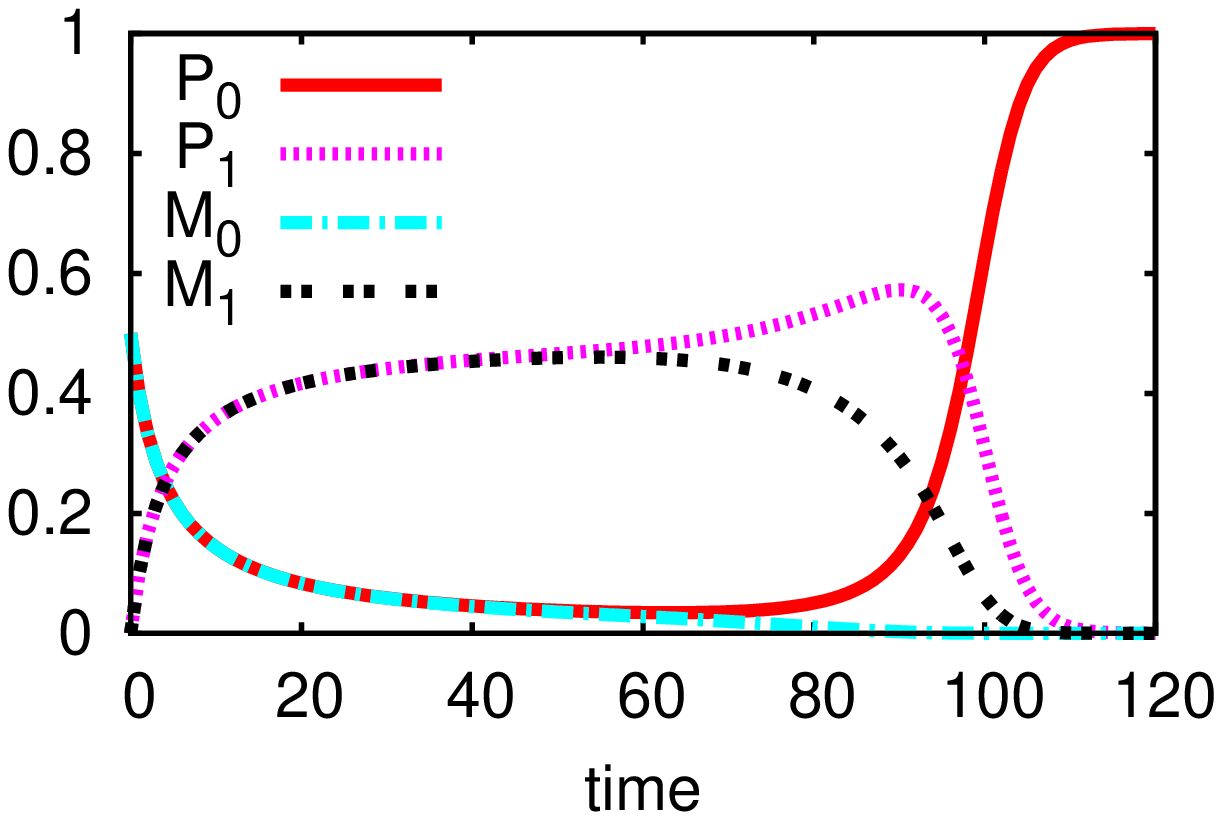}\quad\qquad\includegraphics*[width=0.425\textwidth]{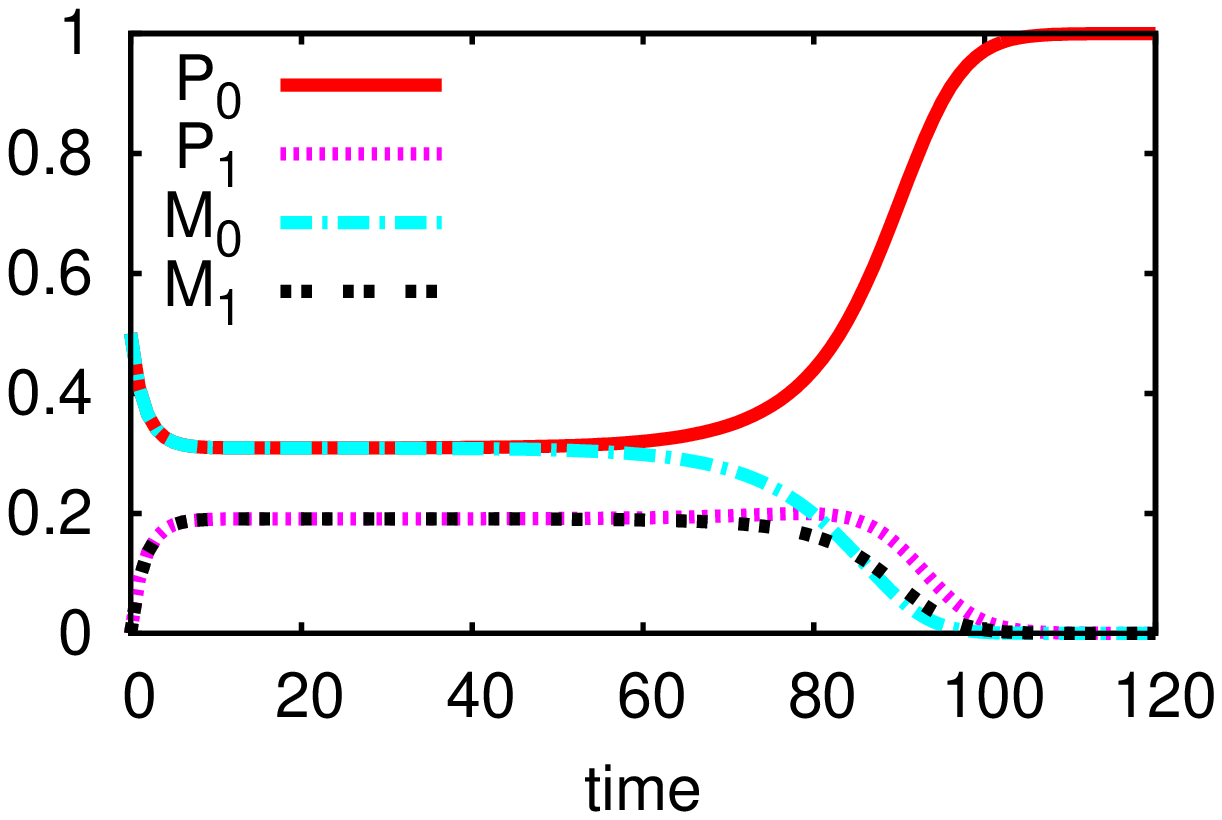}}
\caption{\small Evolution of the densities for the: (left) marginal and
  (right) extremal models with the near-symmetric initial condition
  $P_0=0.50001$, $M_0=0.49999$, and $P_1=M_1=0$.}
  \label{non-symm}
\end{figure}

In the marginal model (Fig.~\ref{non-symm}(a)), the system begins to approach
the point $M_1=P_1=\frac{1}{2}$ algebraically in time, as discussed above.
For a slightly asymmetric initial condition, the densities remain close to
this unstable fixed point for a time that numerical integration shows is of
order $\ln N$.  Ultimately, the system is driven to the fixed point that
corresponds to the initial majority opinion.  For the extremal model,
qualitatively similar behavior occurs, except that in the initial stages of
evolution the system is quickly driven towards the fixed point at
$P_0=M_0=\lambda_+$ and $P_1=\frac{1}{2}-\lambda_+$.  This fixed point is a
saddle node, with one stable and two unstable directions
(Fig.~\ref{non-symm}(b)).  Thus for nearly-symmetric initial conditions, the
densities remain close to this fixed point for a time of the order $\ln N$,
after which the densities are suddenly driven to one of the two stable fixed
points, either $M_0=1$ or $P_0=1$.

\begin{figure}[ht]
\centerline{\includegraphics*[width=0.4\textwidth]{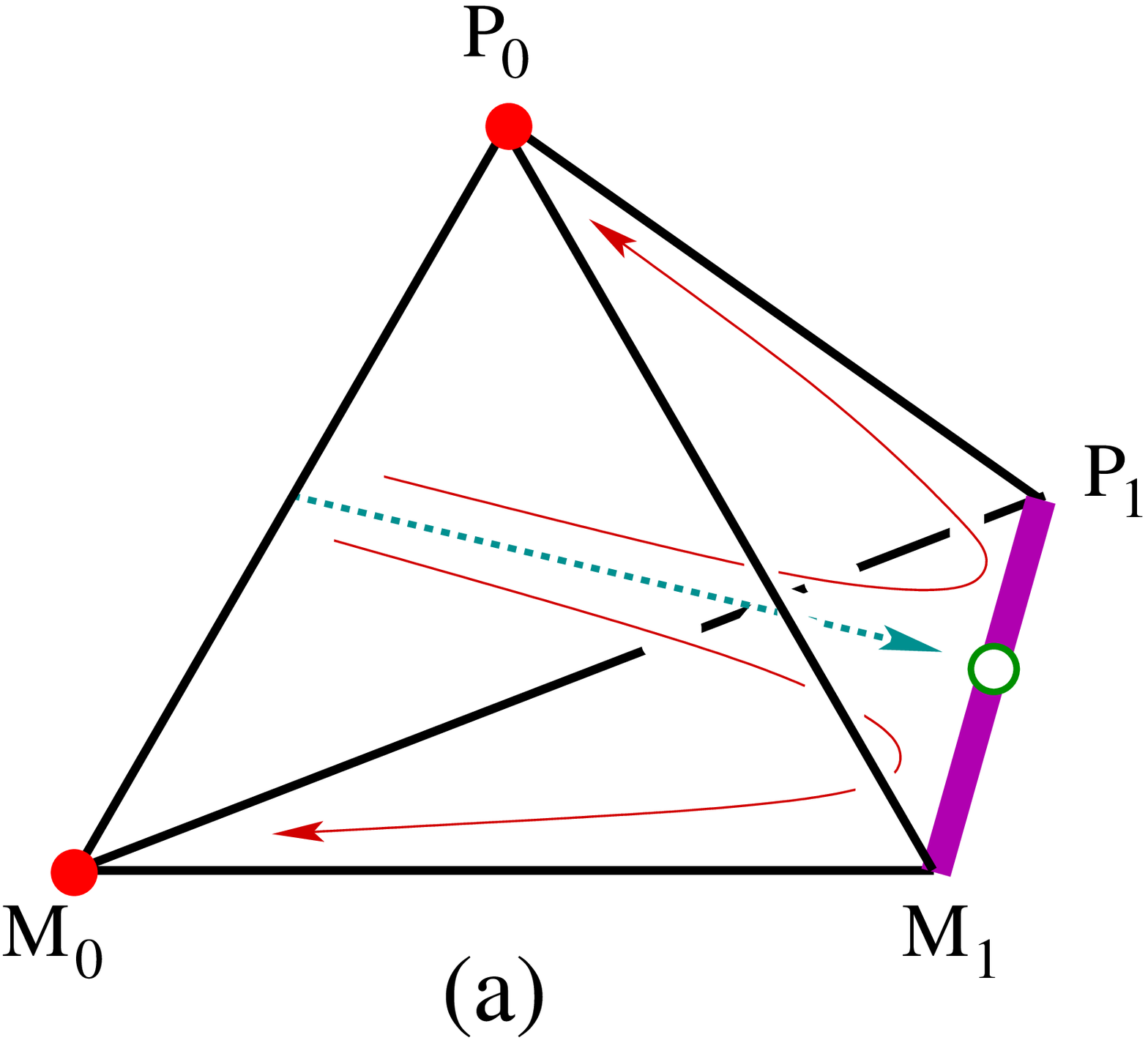}\qquad\qquad\includegraphics*[width=0.4\textwidth]{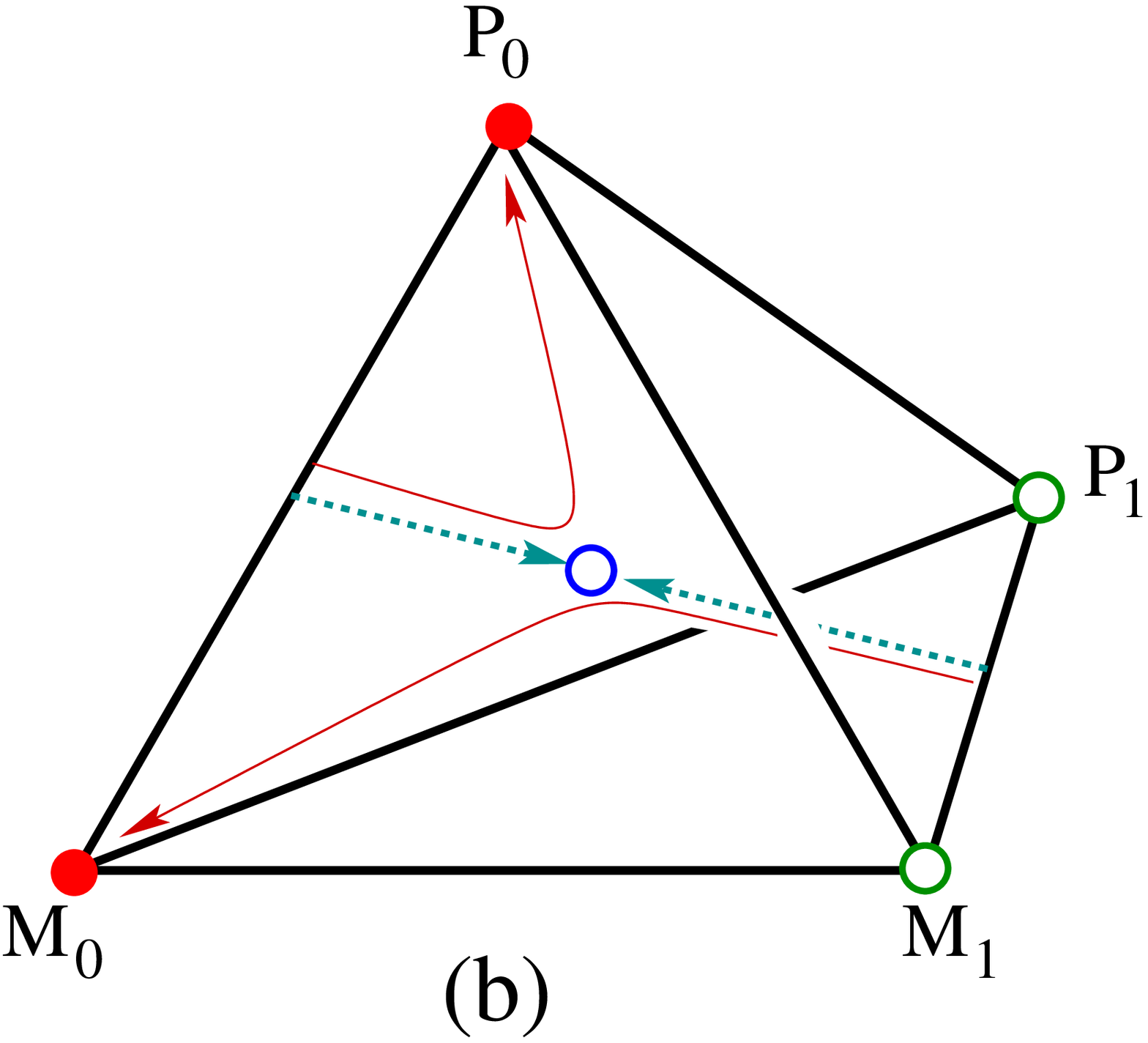}}
\caption{\small Composition tetrahedron for the: (a) marginal and (b)
  extremal models.  Shown in (a) are the consensus fixed points (dots), the
  unstable fixed line (thick), and the symmetry line $P_0=M_0$, $P_1=M_1$
  (dashed arrow) that terminates in a symmetric fixed point (circle).  Shown
  in (b) are the unstable (circle) and stable (dots) fixed points.  For both
  cases, two representative flows that start from nearly symmetric initial
  conditions are shown.}
  \label{tetra-marginal}
\end{figure}

The full state space is the composition tetrahedron, which consists of the
intersection of the set $\{P_0,P_1,M_0,M_1|P_i,M_i\leq 1\}$ with the
normalization constraint plane $P_0+P_1+M_0+M_1=1$
(Fig.~\ref{tetra-marginal}).  Each corner corresponds to a pure system that
is entirely comprised of the labeled species.  For the marginal version,
there are only two stable fixed points at $P_0=1$ and $M_0=1$, corresponding
to consensus of either confident plus voters or confident minus voters.
There is also a fixed line, defined by $P_1+M_1=1$, where the population
consists only of unsure agents.  This fixed line is locally unstable
except at the point $P_1=M_1=\frac{1}{2}$.  Thus if the system starts along
the symmetry line defined by $P_0=M_0$ and $P_1=M_1$, the system flows to the
final state of $P_1=M_1=\frac{1}{2}$.  However, near-symmetric initial states
execute a sharp U-turn and eventually flow to one of the consensus fixed
points $P_0=1$ or $M_0=1$, as illustrated in Fig.~\ref{tetra-marginal}.

For the extremal version, qualitatively similar dynamics arises, except that
instead of a fixed line, there is an unstable fixed point at
$P_0=M_0=\lambda_+$ and $P_1=M_1=\frac{1}{2}-\lambda_+$.  Nearly symmetric
initial states first flow to this unstable fixed point and remain in the
vicinity of this point for a time scale that is of order $\ln N$, after which
the densities quickly flow to the consensus fixed points, either $M_0=1$ or
$P_0=1$.

\section{CONFIDENT VOTING ON LATTICES}
 
We now investigate confident voting dynamics when voters are situated on the
sites of a finite-dimensional lattice of linear dimension $L$ (with $N=L^d$),
with periodic boundary conditions.  For the classic lattice voter model, it
was found that the consensus time $T_N$ asymptotically scales as $N^2$ in one
dimension $d=1$, as $N\ln N$ for $d=2$, and as $N$ for
$d>2$~\cite{Liggett85,K92}.  The presence of the logarithmic factor for $d=2$
and the lack of dimension dependence for $d>2$ shows that the critical
dimension $d_c=2$ for the classic voter model.

The confident voter model has quite different dynamics because the
magnetization is not conserved, except in the symmetric limit $P_0=M_0$ and
$P_1=M_1$, whereas the average magnetization is conserved in the classic
voter model~\cite{Liggett85,K92}.  Here the magnetization is defined as the
difference in the densities of plus and minus voters of any kind.  The
absence of this conservation law leads to an effective surface tension
between domains of plus and minus voters~\cite{DC07}.  Consequently confident
voting is closer in character to the kinetic Ising model with single-spin
flip dynamics at low temperatures rather than to the classic voter model.

\subsection{One Dimension}

\begin{figure}[ht]
\begin{center}
\includegraphics*[width=0.8\textwidth]{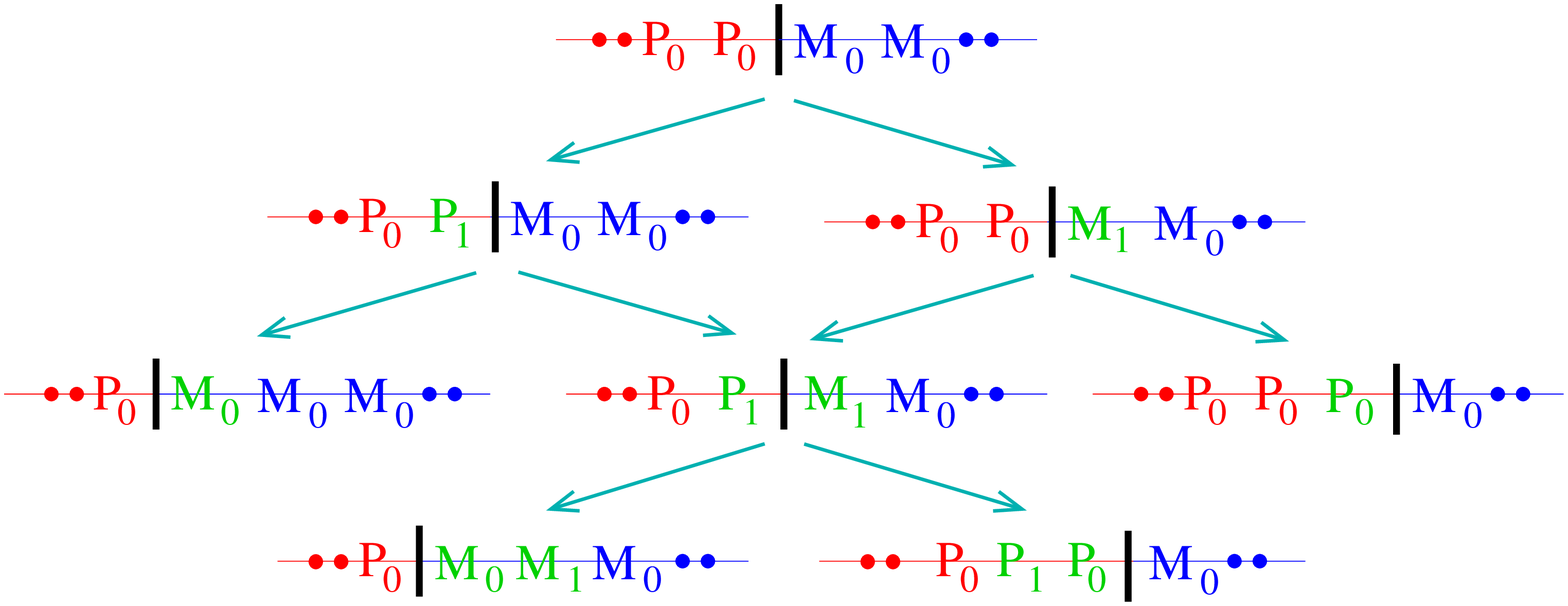}
  \caption{\small First three evolution steps of an interface between a $P_0$
    and $M_0$ domain.  Voters that change their state are shown green.  After
    one more step, a sharp domain wall that is translated by $\pm 1$ lattice
    spacing is re-established. }
  \label{1dwall}
\end{center}
\end{figure}

In the simplest case of one dimension, the agents organize at long times into
domains that are in a single state and the evolution is determined by the
motion of the interface between two dissimilar domains.  Thus we consider the
evolution of a single interface between two semi-infinite domains --- for
example, one in state $P_0$ and the other in state $M_0$.  By enumerating all
possible ways that the voters at the interface can evolve
(Fig.~\ref{1dwall}), we find that the domain wall moves one site to the left
or to the right equiprobably after four time steps.  Thus isolated interfaces
between domains undergo a random walk, but with the domain wall hopping at
one-fourth the rate of a symmetric nearest-neighbor random walk.

Similarly, we determine the fate of two adjacent diffusing domain walls by
studying the evolution of a single voter in state $M_0$ in a sea of $P_0$
voters.  By again enumerating the possible ways these two adjoining
interfaces evolve, we find that the domain walls annihilate with probability
$1/2$ and move apart by one lattice spacing with probability $1/2$.
Additionally, we verified that the distribution of survival times for a
single confident voter in a sea of opposite-opinion voters scales as
$S(t)\equiv t^{-1/2}$, as in the classic voter model.  We also studied the
analogous single-defect initial condition for unsure voters.  In all such
cases, the long-time behavior is essentially the same as in the classic voter
model, albeit with an overall slower time scale.  Finally, we confirmed that
the time to reach consensus starting from an arbitrary initial state scales
quadratically with $N$.  Thus the one-dimensional confident voter model at
long times exhibits the same evolution as the classic voter model, but with a
rescaled time.

\subsection{Two Dimensions}

\begin{figure}[ht]
\centerline{\includegraphics*[width=0.425\textwidth]{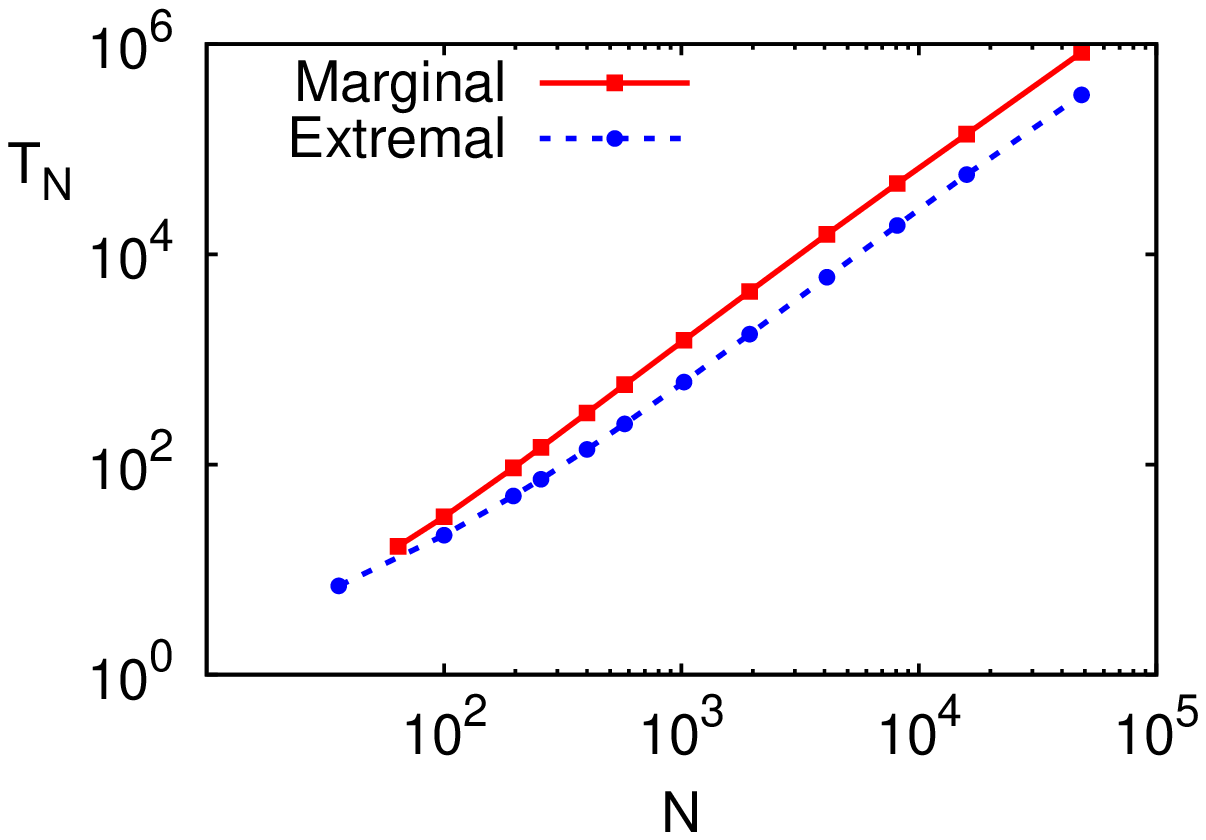}\qquad\includegraphics*[width=0.425\textwidth]{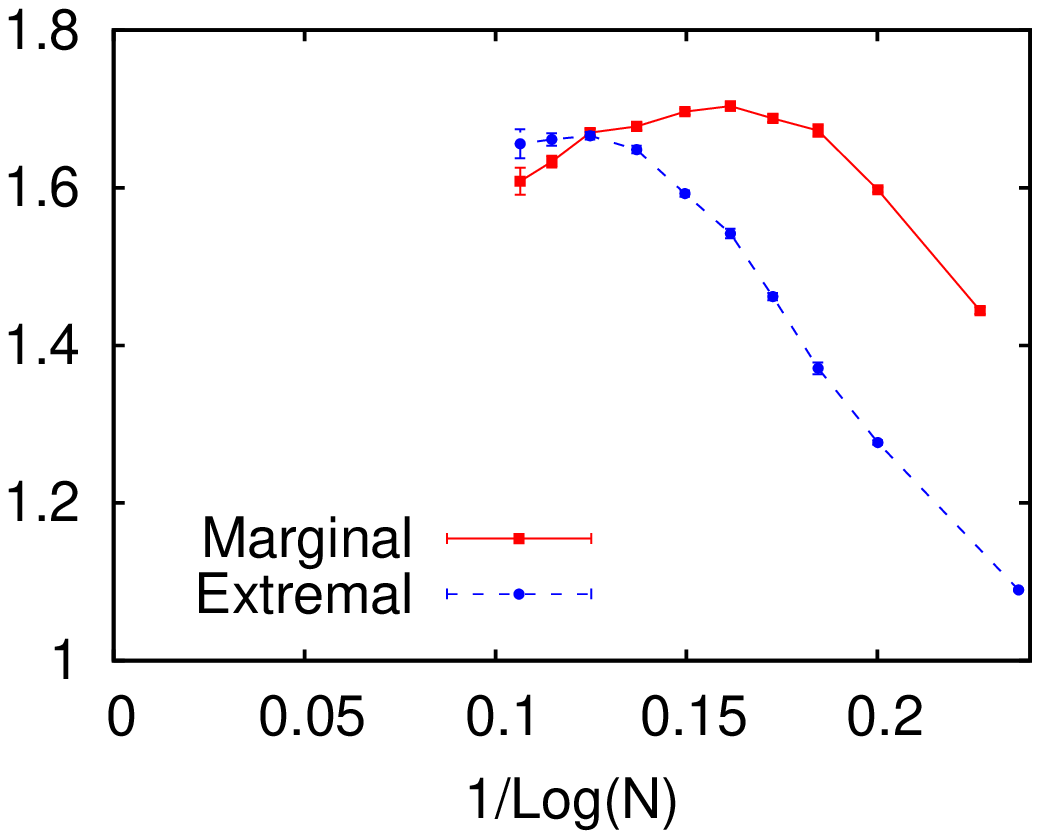}}
\caption{\small (left) Average consensus time $T_N$ on the square lattice as
  a function of $N$.  For both models, the initial number of confident plus
  and minus voters are equal and randomly-distributed in space.  The number
  of realizations for the largest system size is $40,000$. (right) Local
  two-point exponent for the consensus time for the marginal and extremal
  models.  The error bars indicate the statistical uncertainty.}
  \label{T2d}
\end{figure}

In our simulations of confident voting in two dimensions, we typically start
a population with exactly one-half of the voters in the confident plus state and
one-half in the confident minus state, with their locations randomly distributed. 
Periodic boundary conditions are always employed.  For both the marginal and the 
extremal versions of confident voting, $T_N$ appears to grow algebraically in 
$N$, with an exponent that is visually close to $\frac{3}{2}$ (Fig.~\ref{T2d}). 
However, the local two-point slopes in the plot of $\ln T_N$ versus $\ln N$ 
are slowly and non-monotonically varying with $N$ so that it is difficult to 
make a precise estimate of the exponent value.

\begin{figure}[ht]
\centerline{\includegraphics*[width=0.25\textwidth]{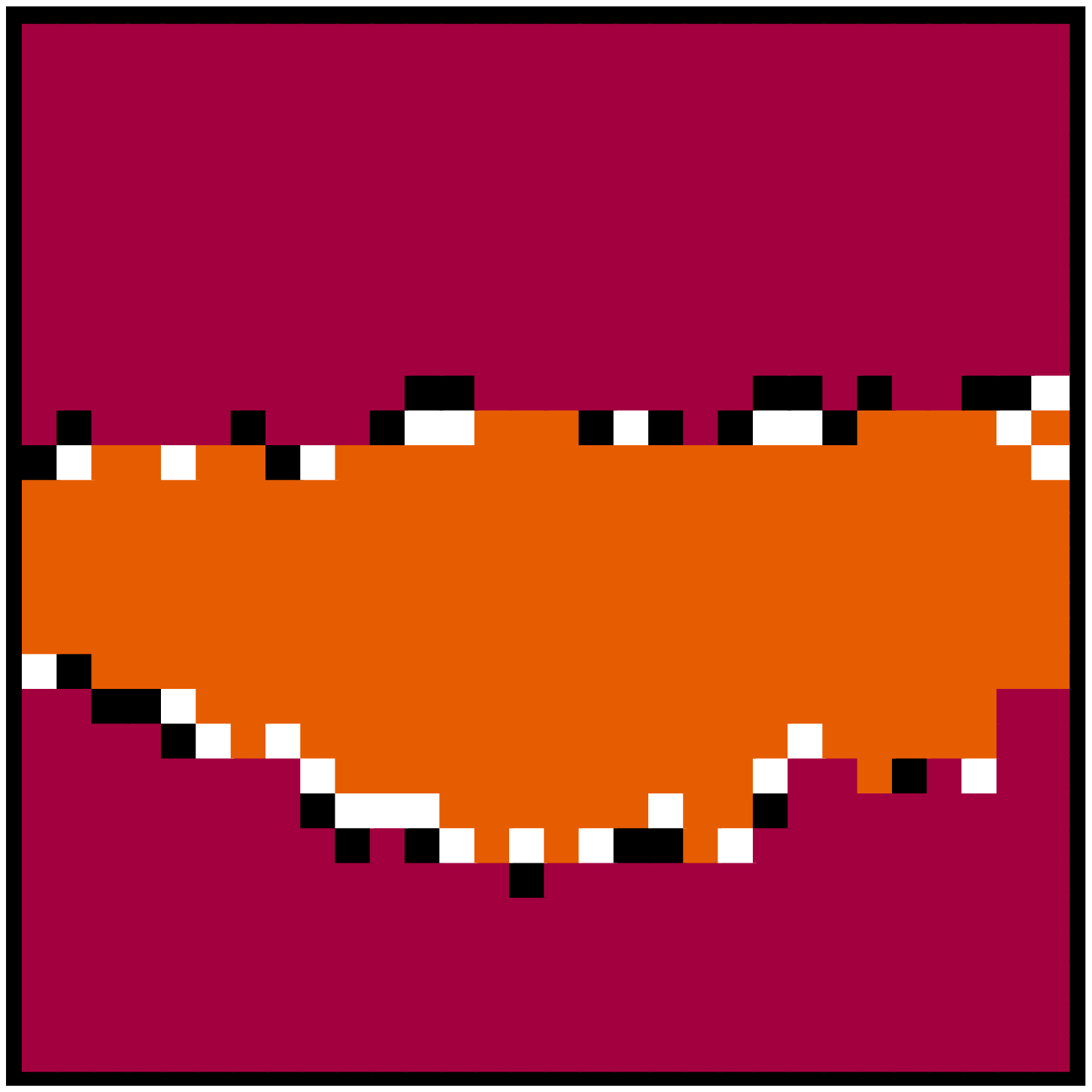}\qquad\qquad\qquad\includegraphics*[width=0.25\textwidth]{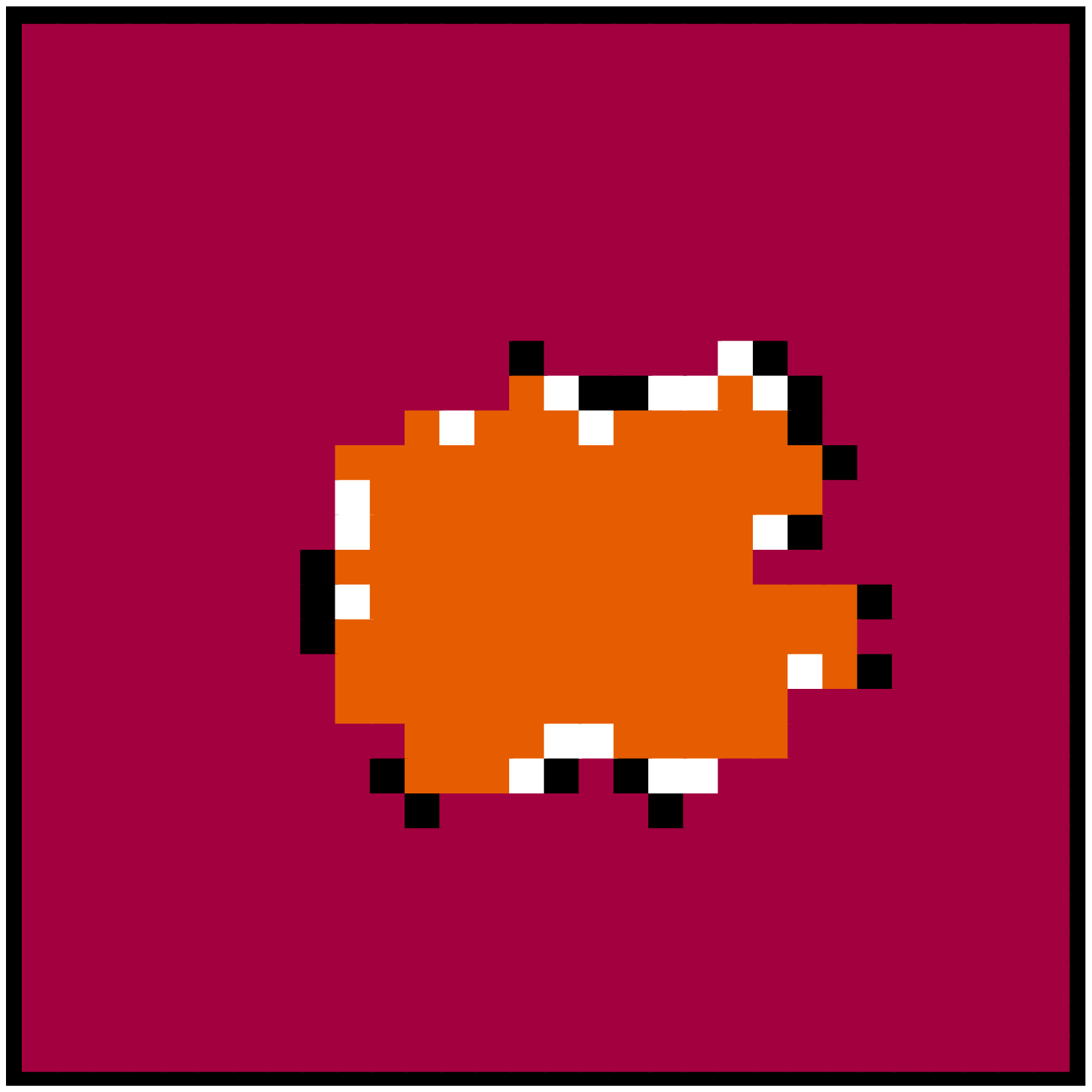}}
\caption{\small Typical configurations of the extremal version of the
  confident voter model on a $30 \times 30$ square lattice that reach either
  a stripe state (left) or an island state (right).  Black and white pixels
  correspond to unsure plus and minus agents; these form a sharp interface
  between domains of confident agents. }
  \label{statePics}
\end{figure}

We argue that this slow approach to asymptotic behavior arises because there
are two different routes by which consensus is achieved.  For random initial
conditions, most realizations reach consensus by domain coarsening, a process
that ends with the formation of a large single-opinion droplet that engulfs
the system.  However, for a substantial fraction of realizations (roughly
38\% for the extremal model and 42\% for the marginal model), voters first
segregate into alternating stripe-like enclaves of plus and minus voters
(Fig.~\ref{statePics}).  This feature is akin to what occurs in the
two-dimensional Ising model with zero-temperature Glauber dynamics, where
roughly one-third of all realizations fall into a stripe state (which happens
to be infinitely long lived at zero temperature~\cite{SKR01a,SKR01b,BKR09}).
A similar condensation into stripe states also occurs in the majority vote
model~\cite{CR05}, the $AB$ model, the naming game~\cite{CBL09}, and now the
confident voter model.  It is striking that this symmetry breaking occurs in
a wide range of non-equilibrium systems for which the underlying dynamics is
symmetric in $x$ and $y$.  It is an open challenge to understand why this
symmetry breaking occurs.

\begin{figure}[ht]
\centerline{\includegraphics*[width=0.5\textwidth]{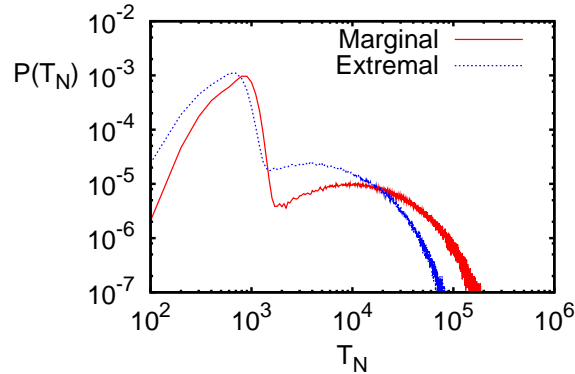}}
\caption{\small Consensus time distribution for a $64 \times 64$ 
  system on a double logarithmic scale.  The initial condition is the same 
  as in Fig.~\ref{T2d} and the data are based on 750,000 realizations. }
  \label{distT2d}
\end{figure}

The existence of these two distinct modes of evolution is reflected in the
probability distribution of consensus times $P(T_N)$ (Fig.~\ref{distT2d}).
Starting from the random but symmetrical initial condition, the distribution
$P(T_N)$ first has a sharp peak at a characteristic time that scales linearly
with $N$, and then a distinct exponential tail whose characteristic decay
time scales as $N^{3/2}$.  The shorter time scale corresponds to the subset
of realizations that reach consensus by conventional coarsening.  For these
realizations, the length scale $\ell$ of the coarsening grows as $\sqrt{t}$.
When this coarsening scale reaches $L$, consensus is achieved.  The consensus
time is thus given by $\ell=L=\sqrt{t}$; since $N\propto L^2$, we have
$T_N\simeq N$.  The longer time scale stems from the subset of realizations
that fall into a stripe state before consensus is eventually reached.

\begin{figure}[ht]
\centerline{\includegraphics*[width=0.425\textwidth]{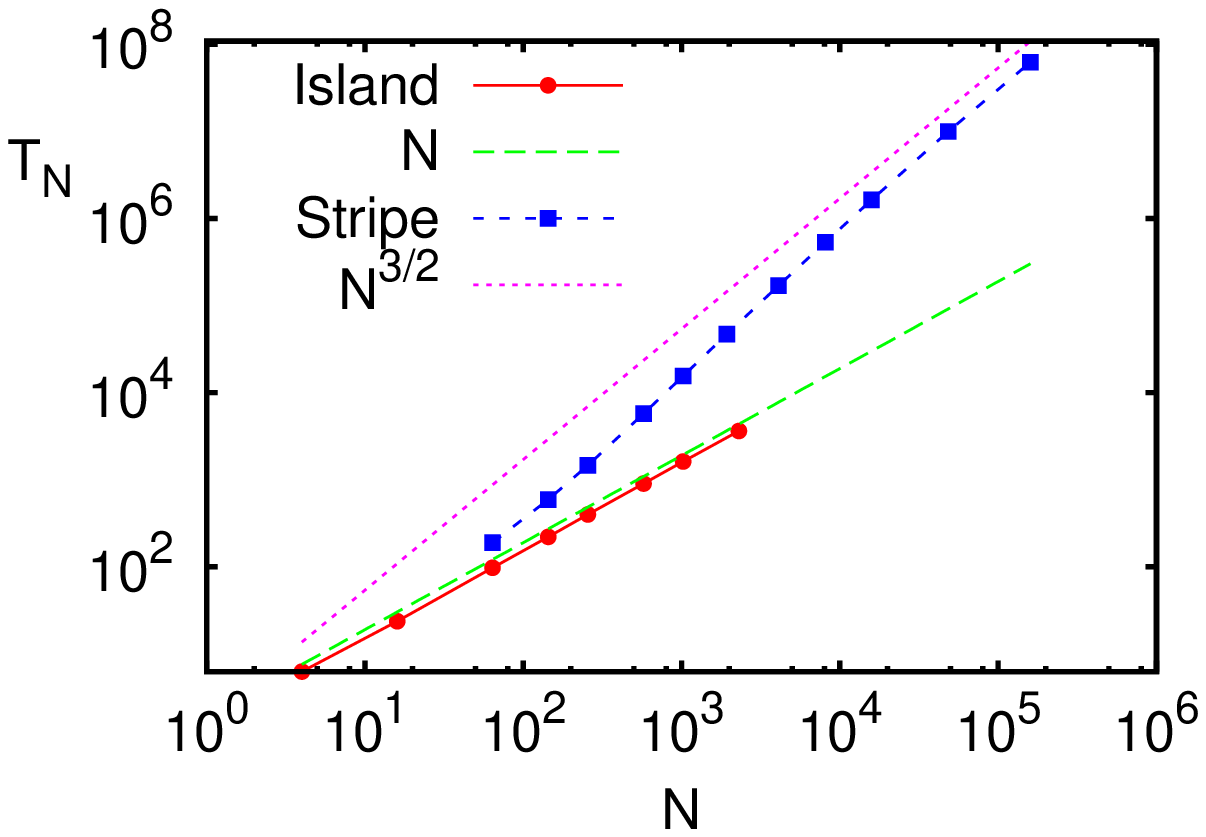}\qquad\includegraphics*[width=0.425\textwidth]{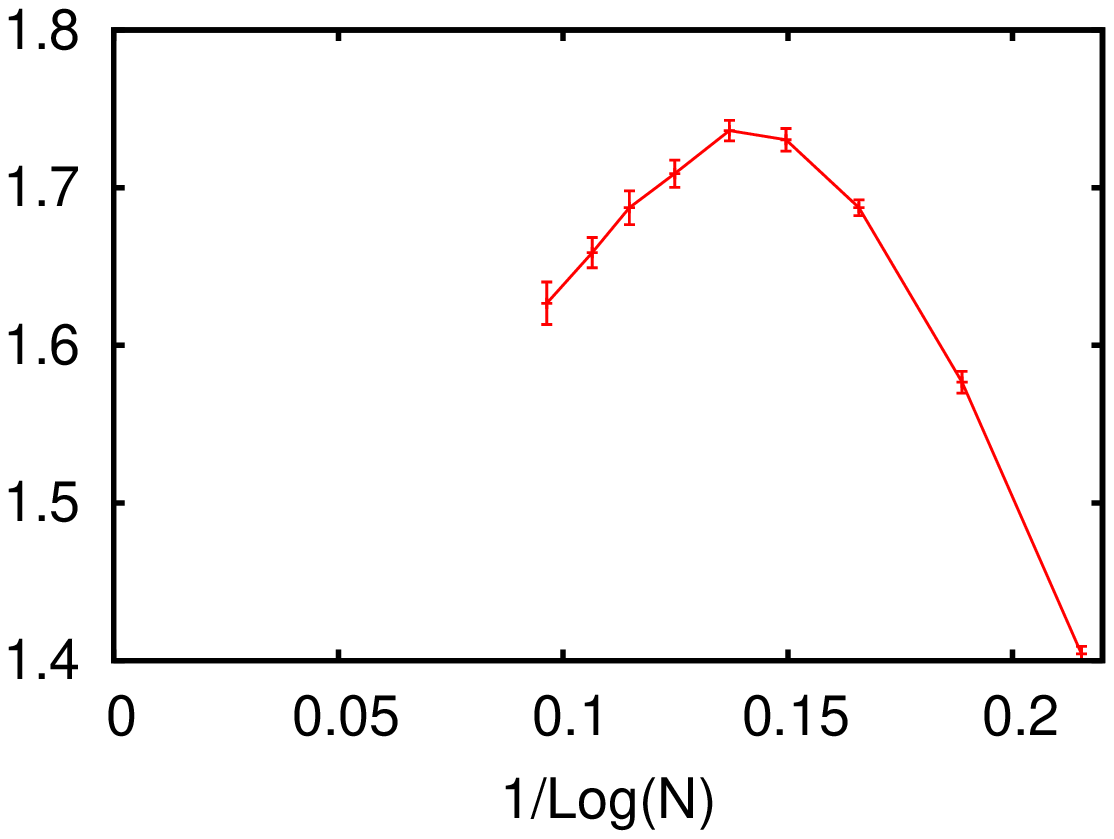}}
\caption{\small (left) Average consensus time for the extremal model on the
  square lattice as a function of the population size for: (i) island and
  (ii) stripe initial configurations.  For the stripe initial condition, the
  data for the largest system size is based on 13,000 realizations.  The
  stripe-state data has been vertically displaced for clarity.  (right) The
  local two-point exponent for $T_N$ for an initial stripe state.  The error
  bars indicate the statistical uncertainty.}
  \label{T2d-combined}
\end{figure}

To help understand the quantitative nature of the approach to consensus via
the two different routes of coarsening and stripe states, we studied the
confident voter model with the initial conditions of: (i) a large circular
single-opinion island and (ii) a stripe state (Fig.~\ref{T2d-combined}).  For
the former, the initial condition is a circular region of radius $r$ that
contains agents in state $M_0$, surrounded by agents in state $P_0$.  For the
latter, agents in state $P_0$ occupy the top half of the system, while the
bottom half is occupied of agents in state $M_0$.  For these two initial
conditions, the consensus time $T_N$ grows as $N$ and as $N^{3/2}$,
respectively (Fig.~\ref{T2d-combined}).  In the latter case, the approach to
asymptotic behavior is both non-monotonic and extremely slow
(Fig.~\ref{T2d-combined}); we do not understand the mechanism responsible for
these anomalies.  These limiting behaviors account for the two time scales
that arise in the distribution of consensus times for a system with a random,
symmetric initial condition.

Although the confident voter model has an appreciable probability of falling
into a stripe state, such a state is not stable because the interface between
the domains can diffuse.  When the two interfaces of a stripe diffuse by a
distance that is of the order of their separation, one stripe is cut in two
and resulting droplet geometry quickly evolves to consensus.  We estimate the
time for two such interfaces to meet by following essentially the same
argument as that developed for the majority vote model~\cite{CR05}.  For a
flat interface, every site on the interface can change its opinion.  Such an
opinion change moves the local position of the interface by $\pm 1$.  For a
smooth interface of length $L$, there will therefore be of the order of
$L\pm\sqrt{L}$ opinion change events of plus to minus and vice versa.  Thus
the net change in the number of agents of a given opinion is of the order of
$\pm\sqrt{L}$.  Consequently, the average position of the interface moves by
$\sqrt{L}/L=1/\sqrt{L}$.  Correspondingly the diffusion coefficient $D_L$ of
the interface scales as $1/L$.  The time for two such interfaces that are
separated by a distance of the order of $L$ to meet therefore scales as
$L^2/D_L\sim L^3\sim N^{3/2}$.

In a $d$-dimensional system, the analog of two-stripe state is a two-slab
state with a $(d-1)$-dimensional interface separating the slabs.  Now the
same argument as that give above leads to $N^{(d+1)/d}$ as the time scale for
two initially flat interfaces to meet.  According to this approach, the
consensus time scales linearly with $N$ in the limit of $d\to\infty$, a limit
that one normally associates with the mean-field limit.  However, the rate
equation approach gives a consensus time that grows as $\ln N$.  We do not
know how to resolve this dichotomy.

\section{Summary}

We introduced the notion of individual confidence in the context of the voter
model.  Our model is based on recent social experiments that point to the
importance of multiple reinforcing inputs as an important influence for
adopting a new opinion or behavior~\cite{C10}.  We studied two variants of
confident voting in which an agent who has just switched opinion will be
either have confidence in the new opinion --- the extremal model --- or be
unsure of the new opinion --- the marginal model.  In the mean-field limit, a
nearly symmetric system quickly evolves to an intermediate metastable state
before finally reaching a consensus in one of the confident opinion states.
This intermediate state is reached in a time of the order of one, while the
time to reach consensus scales as $\ln N$.

On a two-dimensional lattice, a substantial fraction of all realizations of a
random initial condition reach a long-lived stripe state before ultimate
consensus is reached.  This phenomenon appears ubiquitously in related
opinion and spin-dynamics models~\cite{CR05,SKR01a,SKR01b,BKR09}) and an
understanding of what underlies this dynamical symmetry-breaking is still
lacking.  An important consequence of the stripe states is that there are two
independent times that describe the approach to consensus.  The shorter time,
which scales linearly with $N$, corresponds to realizations that reach
consensus by domain coarsening.  The longer time corresponds to realizations
that get stuck in a metastable stripe state before ultimately reaching
consensus.

An unexpected feature of confident voting is that the behavior in two
dimensions, where the consensus time $T_N$ varies as a power law in $N$, is
drastically different than that of the mean-field limit, where $T_N$ varies
logarithmically with $N$.  In contrast, in the classic voter model, $T_N\sim
N\ln N$ in two dimensions, whereas the mean-field behavior is $T_N\sim N$.
This dichotomy suggests that confident voting on the complete graph does not
correspond to the limiting behavior of confident voting on a high-dimensional
hypercubic lattice.  Moreover the argument that $T_N$ on a $d$-dimensional
hypercubic lattice scales as $N^{(d+1)/d}$ suggests that the upper critical
dimension for confident voting is infinite.

\ack We gratefully acknowledge financial support from NSF Grant No.\ DMR-0906504.

\def\newblock{\hskip .11em plus .33em minus .07em}
%\bibliographystyle{apsrev}
%\bibliography{refs}

\end{document}